\documentclass[manuscript]{aastex6}

\usepackage{natbib}
\usepackage{graphicx}

\usepackage{amsmath}

\usepackage{gensymb}      

\usepackage[T1]{fontenc}  
\usepackage{textcomp}     
\usepackage{mathptmx}     

\usepackage{multirow}

\newcommand{\kepler}[0]{\emph{Kepler}}
\newcommand{\lsst}[0]{\emph{LSST}}
\newcommand{\kebc}[0]{KEBC}
\newcommand{\maf}[0]{MAF}
\newcommand{\vartools}[0]{VARTOOLS}
\newcommand{\nasa}[0]{NASA}
\newcommand{\aov}[0]{AoV}
\newcommand{\opsim}[0]{OpSim}
\newcommand{\galaxia}[0]{\emph{Galaxia}}

\newcommand{\pratio}[0]{P_r}
\newcommand{\plsst}[0]{P_{LSST}}
\newcommand{\pkebc}[0]{P_{KEBC}}

\newcommand{\lfov}[0]{9.6~deg$^{2}$}
\newcommand{\kfov}[0]{115~deg$^{2}$}
\newcommand{\kra}[0]{19h22m40s}
\newcommand{\kdec}[0]{+44\degree{}30\textquotesingle{}00\textquotedbl}

\definecolor{codebackground}{rgb}{0.8,0.8,0.8}
\newcommand{\code}[1]{\colorbox{codebackground}{\textbf{\texttt{#1}}}}

\begin{document}
    \title{Initial Estimates on the Performance of the \lsst{} on the Detection of Eclipsing Binaries}

    \author{
        Mark Wells\altaffilmark{1,2},
        Andrej Pr\u{s}a\altaffilmark{2},
        Lynne Jones\altaffilmark{3},
        Peter Yoachim \altaffilmark{3}}

        \altaffiltext{1}{Department of Astronomy \& Astrophysics, The Pennsylvania State University, University Park, PA 16802, USA}
        \altaffiltext{2}{Department of Astronomy \& Astrophysics, Villanova University, 800 E. Lancaster Avenue, Villanova, PA 19085, USA}
        \altaffiltext{3}{Department of Astronomy, University of Washington, 3910 15th Ave NE, Seattle, WA 98195, USA}

    \shorttitle{\lsst{} Detection of Eclipsing Binaries}
    \shortauthors{Wells}

\begin{abstract}
    In this work we quantify the performance of \lsst{} on the detection of eclipsing binaries.
    We use \kepler{} observed binaries to create a large sample of simulated pseudo-\lsst{} binary light curves.
    From these light curves, we attempt to recover the known binary signal.
    The success rate of period recovery from the pseudo-\lsst{} light curves is indicative of \lsst{}'s expected performance.
    Using an off-the-shelf Analysis of Variance (AoV) routine, we successfully recover 71\% of the targets in our sample.
    We examine how the binary period impacts recovery success and see that for periods longer than 10~days the chance of successful binary recovery drops below 50\%.
\end{abstract}

\section{Introduction}
    \subsection{\lsst{} and \kepler{}}
    The Large Synoptic Survey Telescope (\lsst{}) is unlike any previous telescope survey.
    Capable of imaging the entire sky approximately four times in a single week, the amount of data it will generate is substantial: the 3.2-gigapixel camera will take approximately 1000~pairs of exposures generating roughly 15 Terabytes of data each night.
    Currently under construction in Chile, the ten-year survey is expected to begin in 2022.
    The \lsst{} fast cadence makes \lsst{} appealing for the study eclipsing binaries, especially at the faint end where such an extensive survey was hereunto not possible.

    The value of eclipsing binaries cannot be overstated.
    Aside from being exquisite distance estimators \citep{Ribas2005}, binary stars allow for direct, model-independent,  measures of stellar mass \citep{Torres2010}.
    The eclipses provide the relative stellar radii, the inclination of the system, and orbital characteristics such as the eccentricity and the argument of periastron.
    Even without radial velocities, relative radius estimates can be made for systems.

    \lsst{} will allow us to perform large scale binary population studies.
    One aspect that \lsst{} will be uniquely equipped to study is binary rates as a function of spectral type, especially at the low mass end.
    At the high mass end, at least 75\% of O-type stars form in binary or multiple systems \citep{Mason1998,Mason2009}.
    Lower mass stars are less likely to be found in binaries:
    70\% of B-types and A-types \citep{Shatsky2002,Kobulnicky2007,Kouwenhoven2007},
    roughly 50\% of solar-types \citep{Ragavon2010},
    and M-Dwarfs estimates range between 11\% \citep{Reid1997} up to 34-42\% \citep{Henry1990,Fischer1992}.
    For the least massive stars and brown dwarfs, it is estimated that only 10\%-30\% of stars exist in multi-star systems \citep{Burgasser2003,Siegler2005,Allen2007,Maxted2008,Joergens2008}.

    More massive stars form in massive gas clouds which tend to form multiple stars closer together, leading to the formation of systems with multiple companions.
    As can be seen from the previously stated multiplicity rates, the less massive a star, the smaller the multiplicity rate.
    This, compounded with their faintness, makes low mass binaries difficult to study.
    \lsst{} will be able to survey faint M-M binary pairs thanks to its large \'{e}tendue.
    
    In order to estimate \lsst{}'s performance on observing eclipsing binary systems we look to \nasa{}'s \kepler{} mission \citep{Borucki2010}.
    This mission was designed to discover Earth-size exoplanets in or near the habitable zones of their host stars using the transit method of planet detection.
    This method requires a convenient orientation of the system being observed: the plane of the planet's orbit must be aligned with the line of sight such that the planet will pass in front of the parent star and block some of its light.
    The \kepler{} telescope was launched on March~7, 2009 into an Earth-trailing orbit about the Sun to achieve a high pointing stability.
    \kepler{} produced observations of unprecedented quality in terms of sensitivity and duration.
    It created a superb sample of binaries due to long cadence of approximately 30~minutes, short cadence of approximately 1~minute, mission lifetime of 4~years, and observations of approximately 200,000 targets.
    The \kepler{} Eclipsing Binary Catalog (\kebc{}) \citep{kebc1,kebc2,kebc3,kebc4} provides a robust and vetted list of 2876~binaries as of July~1,~2016.
    The list of the catalog fields and a short description of each is given in Table~\ref{tab:kebcCols}.
    \floattable
    \begin{deluxetable}{cl}
        \tablecolumns{2}
        \tablecaption{
            Column names given in the \kebc{} with a short description.
            \label{tab:kebcCols}}
        \tablehead{
            \colhead{Catalog Column Name} & \colhead{Description}}
        \startdata
            KIC          & the \kepler{} input catalog identification number \\
            period       & the period of the binary \\
            period error & the associated uncertainty in the period \\
            bjd0         & time of primary eclipse \\
            bjd0 error   & uncertainty in bjd0 \\
            GLon         & galactic longitude \\
            GLat         & galactic latitude \\
            kmag         & \kepler{} magnitude \\
            LC data      & long cadence \kepler{} archived data \\
            SC data      & short cadence \kepler{} archived data \\
        \enddata
    \end{deluxetable}

\subsection{Simulation Overview}
    The \kebc{} sources served as templates which we used to generate light curves that resemble what \lsst{} will observe.
    First, we considered a region of the sky for which we have a fairly complete sampling, namely the \kepler{} field.
    We then use this field to create a sample field in the Southern hemisphere along the same galactic latitude, in order to preserve the stellar number density.
    \lsst{}'s performance on this equivalent field is evaluated by simulating observations of these sources according to the \lsst{} cadence.

    The \lsst{} cadence will be non-uniform and determined by a multitude of inputs.
    Weather conditions, slew distance, and science goals are all considered when determining which part of the sky to observe next.
    The next field \lsst{} observes is determined by optimization of all of these conditions.
    Until \lsst{} is actually in operation, we can only simulate its observing cadence.
    The \lsst{} science team has developed the \lsst{} Operations Simulator (\opsim{}, \citealt{opsim}) which will enable the simulation of the \lsst{} cadence.
    The cadence of the light curves was extracted from \opsim{} using the Metrics Analysis Framework (\maf{}, \citealt{maf}).
	\opsim{} and \maf{} are developed as part of the overall \lsst{} simulations effort \citep{Connolly2014}.
    Finally, the pseudo-\lsst{} light curves were interpolated from the \kebc{} light curves using the \lsst{} cadence.
    We neglect targets with multiple binary signals (i.e. systems with more than two members) in this current treatment.
    
    We next attempt to recover the binary signal from the newly created light curves.
    For a target to be considered recovered as a binary, the strongest period will need to be a harmonic (integer multiple or fraction) of the known \kebc{} period.

    This work serves as a proof of concept and a first step that will lead to a more sophisticated analysis including the production of an eclipsing binary science metric for use in the construction of \lsst{}'s observing cadence.

\newpage
\section{Simulated Pseudo-\lsst{} Light Curves}
    \label{sec:simlsst}
\begin{figure}[!tbp]
        \centering
        \begin{tabular}{cc}
            \includegraphics[width=0.5\textwidth]{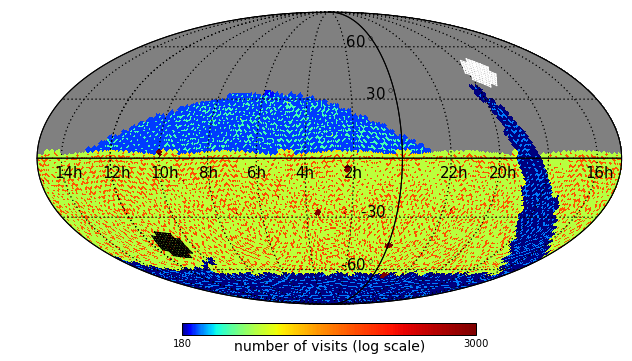} &
            \includegraphics[width=0.5\textwidth]{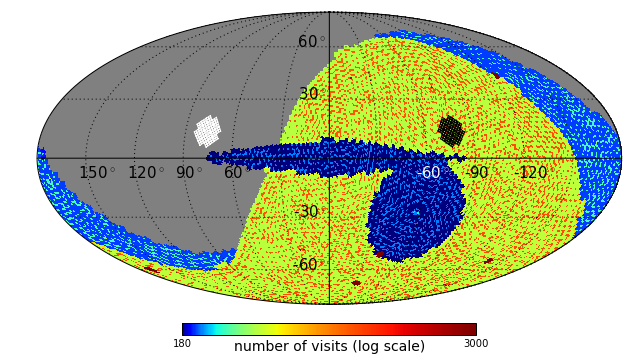}
        \end{tabular}
        \caption{
            Plots of the simulated number of visits \lsst{} will make over its planned 10~year mission lifetime.
            A log scale was chosen to emphasize the different observing strategies \lsst{} will employ.
            \textbf{Left}: Plotted in equatorial coordinates with the \kepler{} field in white and the translated field in black.
            \textbf{Right}: Plotted in galactic coordinates with the \kepler{} field in white and the translated field in black.
            \label{fig:kepoverlay}
        }
    \end{figure}

    A modification was made to the \kebc{} light curves to make them appropriate for this work.
    Because \kepler{} observed a fixed region in the Northern hemisphere (RA \kra{}, Dec \kdec{}), it was necessary to translate the dataset to the southern sky and into the viewing range of \lsst{}.
    To achieve this, the galactic longitude ($l$) was inverted for each \kebc{} target such that $l \rightarrow - l$.
    By applying this transformation we are able to preserve the galactic latitude of our sample, which builds on the assumption that the number density of stars is, in general, uniform along galactic latitude.
    The result of this operation is depicted in Figure~\ref{fig:kepoverlay}.
    Comparing the \kepler{} field coverage, \kfov{}, to the \lsst{} field of view per exposure, \lfov{}, shows that roughly 11 \lsst{} fields would cover the \kepler{} field.
    This also demonstrates that the \kepler{} field is a large enough patch of sky to provide multiple pointings and hence will allow separate cadences to be generated for samples.

\section{\lsst{} Simulations}
    \label{sec:lsstsim}
    The actual \lsst{} observing cadence will be determined as the mission progresses, with changes occurring on a nightly basis due to factors such as changing weather conditions.
    Therefore, the ten-year observing strategy must be simulated.

    \opsim{} is a tool that generates a realistic pointing history for \lsst{} \citep{opsim}.
    \opsim{} takes many factors into account: realistic weather conditions such as clouds, and their effect on seeing; downtime due to scheduled as well as unscheduled maintenance; a high-fidelity model of the telescope behavior including slew time and filter changes; and a scheduler that selects desired observations for each pointing.
    \opsim{} outputs information about each observation that was simulated (e.g., epoch, sky position, seeing, sky brightness).

    The current reference baseline run (\emph{minion\_1016}) was processed with \opsim{} and the output was then analyzed using \maf{}.
    For this treatment we use \maf{} to retrieve the exposure times for our binary targets.
    \maf{} allows for metrics to be defined that measure quantities of interest from \opsim{}'s simulation of the pointing history.


\subsection{Light Curve Creation}
    A \code{UserPointSlicer} was used to obtain the exposure times.
    This slicer accepts a list of RA and Dec coordinates as input and returns a nested list of the exposure times for each point defined by the user.

    The time stamps for each \kebc{} binary were phase-folded over their respective period.
    The flux values were not simulated, instead we adopted the interpolated flux values from the \kepler{} phase-folded light curve.
    This flux interpolation is effectively sampling the \kepler{} light curve with the \lsst{} cadence.
    The resulting pseudo-\lsst{} phase-folded light curve is then unfolded back into time domain, producing an \lsst{} light curve.
    Flux uncertainties were similarly interpolated and adopted.

\begin{figure}[!tbp]
        \centering
        \begin{tabular}{cc}
            \includegraphics[width=0.5\textwidth]{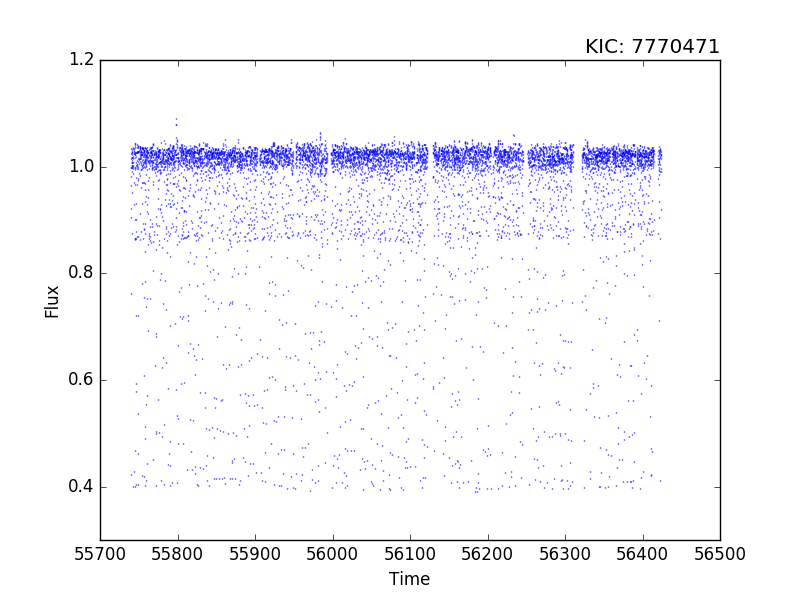} &
            \includegraphics[width=0.5\textwidth]{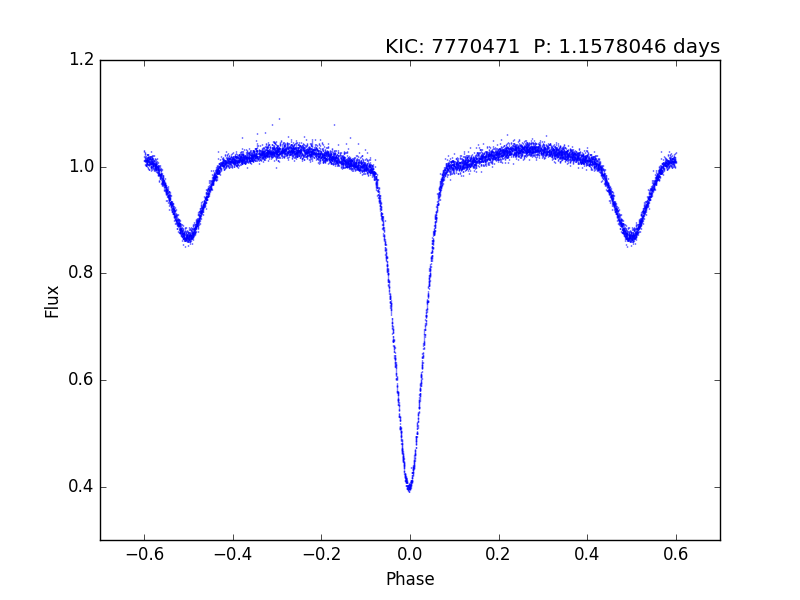} \\
            \includegraphics[width=0.5\textwidth]{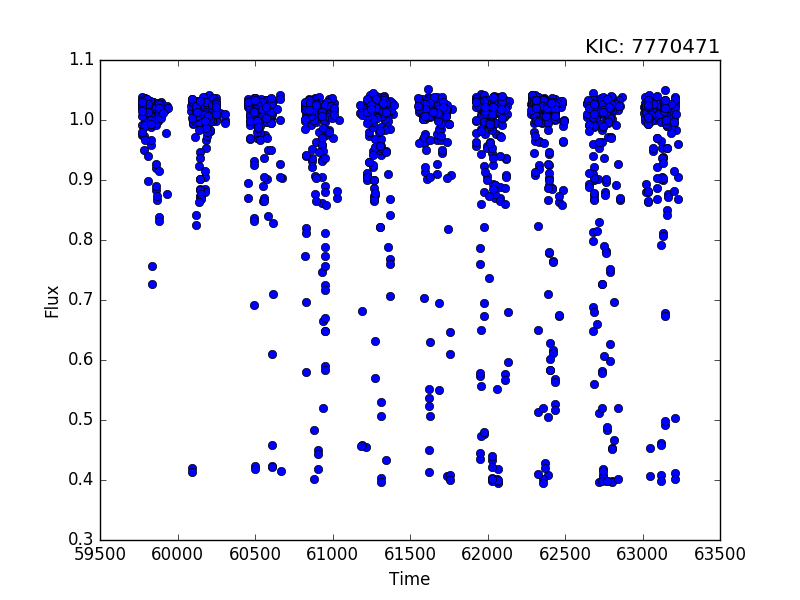} &
            \includegraphics[width=0.5\textwidth]{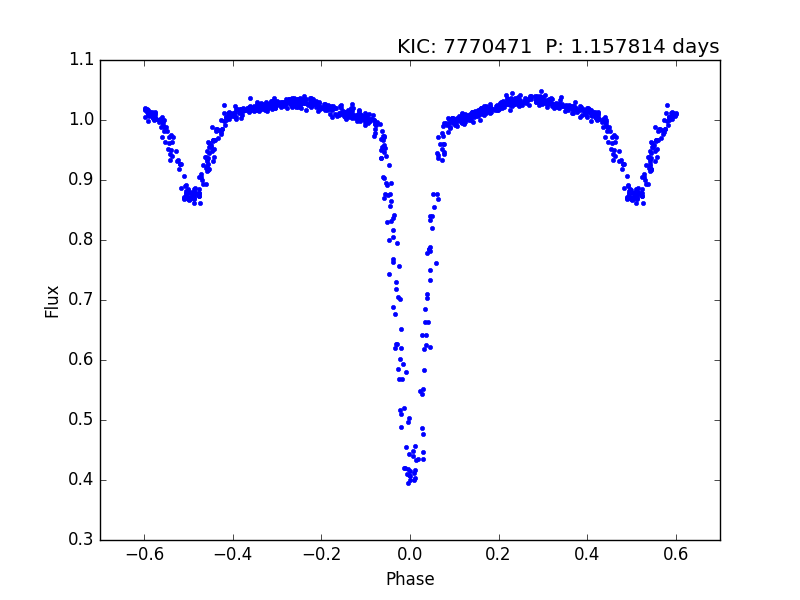}
        \end{tabular}
        \caption{
            \textbf{Top Left}: Original \kepler{} light curve.
            \textbf{Top Right}: The light curve phase-folded over the \kebc{} period.
            \textbf{Bottom Left}: \lsst{} simulated light curve.
            \textbf{Bottom Right}: The light curve phase-folded over the \aov{} period.
        }
        \label{fig:perfigs}
    \end{figure}

\begin{figure}[!tbp]
        \includegraphics[width=\textwidth]{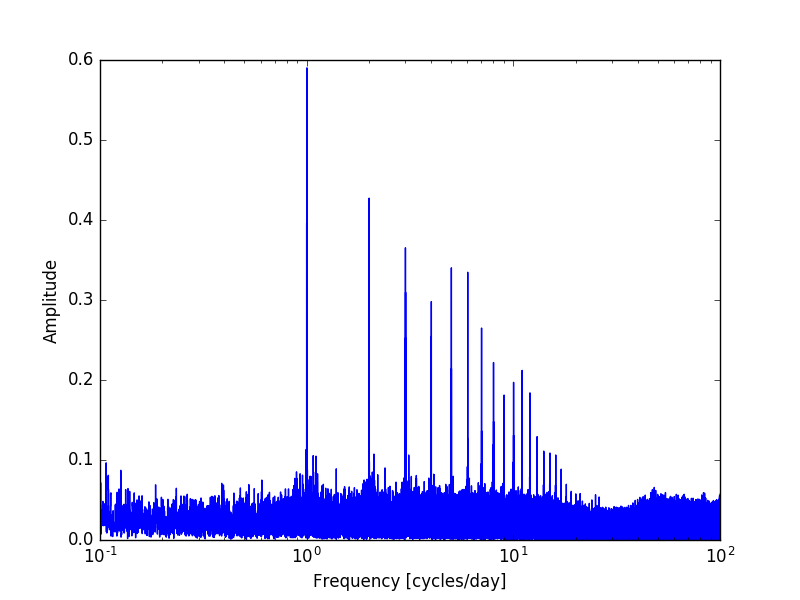}
        \caption{
            The window function of a constant source as observed by \lsst{}.
            The diurnal cycle can be clearly seen.
        }
        \label{fig:winfun}
    \end{figure}

    These light curves are \lsst{}-like, meaning that the time domains are representative of what one would expect.
    However, it is worth pointing out the differences between these missions.
    The \kepler{} instrument used single broad band photometry while \lsst{} will have 6 passbands, \emph{ugrizy}, preferentially making observations in the \emph{r}- and \emph{i}-bands.
    The magnitude limit of each mission is also vastly different.
    \lsst{}'s main survey will observe down to a typical \emph{r} magnitude of 24.5 \citep{ivezic2008}, while \kepler{} had a breakdown limit of approximately 19\textsuperscript{th} magnitude \citep{kephandbook}.
    We use flux values which are derived from \kepler{} and no conversion factors or corrections were made beyond the detrending which was applied to the original light curves.
    The flux uncertainties are certainly problematic as they are reflective of \kepler{}'s systematics and data acquisition.
    Therefore, these light curves can at this time provide only qualitative insight, quantifying only how the cadence of \lsst{} will impact the detection of binary stars.
    Figure~\ref{fig:perfigs} shows a \kebc{} light curve in the top panels and the simulated \lsst{} light curve in the bottom panels.
    For reference we present a plot of a typical window function for \lsst{} in Figure~\ref{fig:winfun}.
    The diurnal cycle is obvious and will be present in all of the data.
    Red noise (cf. \citet{Pont2006}) is expected to be relatively low for \lsst{}.
    This expectation is based upon the irregularity of the \lsst{} cadence as red-noise is most notable in back-to-back time-series.
    However, it should be noted that the \opsim{} simulator does not provide enough fidelity at this time to determine the impact and influence of red noise.

    \lsst{} has a limiting magnitude of 15.8 for a 15-second $r$ band exposure \citep{LSSTScienceBook}.
	Therefore, we look at what precision \lsst{} could achieve while looking at a $16^\textnormal{th}$~mag point source in $r$.
    \citet{ivezic2008} gives the random photometric error for a single \lsst{} visit, $\sigma_\textnormal{rand}$, of a point source of a certain magnitude, $m$, as
    \begin{align*}
        \sigma^2_\textnormal{rand} = (0.04-\gamma)x+\gamma x^2\hspace{5pt}(\textnormal{mag}^2)
    \end{align*}
    with $x \equiv 10^{0.4(m-m_5)}$.
    The $\gamma$ term collects a number of factors including sky brightness and readout noise among others, while $m_5$ is the $5\sigma$ depth for a specific band.
    \citet{ivezic2008} gives a value of 0.039 for $\gamma$.
    The $5\sigma$ depth depends upon the sky conditions and telescope orientation.
    For a typical point source near zenith, with the expected median sky brightness, \citet{ivezic2008} gives a value of 24.43 for $m_5$.
    For a $16^\textnormal{th}$ mag point source we have $\sigma_\textnormal{rand}=657$~ppm.
    
    Turning our attention to \kepler{}, we look at the Combined Differential Photometric Precision (CDPP).
    This metric, used in the transiting planet search (TPS) pipeline \citep{Jenkins2010, Tenebaum2012} is a measure of the white noise of a time series as seen by a transit.
    \citet{Christiansen2012} gives a median CDPP value of 139.1~ppm for a $15^\textnormal{th}$~mag dwarf with a 6~hour transit.
    Comparing the expected value of the \lsst{} photometric precision that we calculated to this value we find that this is roughly 5 times larger.
    To simulate this, we injected a per-point Gaussian noise at the $5\sigma$ level into the simulated light curves.

\newpage
\section{Period Detection}
    The pseudo-\lsst{} light curves are passed to the Analysis of Variance (AoV) tool \citep{aov1,aov2} as implemented in \vartools{} \citep{vartools}.
    This command line utility provides an assortment of tools and methods for analyzing light curves.
    For each simulated light curve the method performs a period search by phase-binning the light curves and calculating the $\theta\_aov$ statistic over a specified period range.
    The period, the $\theta\_aov$, the signal-to-noise ratio, and the formal false alarm probability are computed.

    For our treatment we only look at the period with the highest $\theta\_aov$ which we denote as $\plsst{}$.
    Every target will have a computed $\plsst{}$ but we need to verify that the strongest signal corresponds to the known binary period.
    Next, $\plsst{}$ is compared to the known period from the catalog, $\pkebc{}$.
    In order to evaluate the effectiveness, we compute a period ratio, $\pratio{}$, defined as:
    \begin{align*}
        \pratio{} \equiv \frac{\plsst}{\pkebc}.
    \end{align*}

\begin{figure}[!tbp]
        \centering
        \begin{tabular}{cc}
            \includegraphics[width=0.5\textwidth]{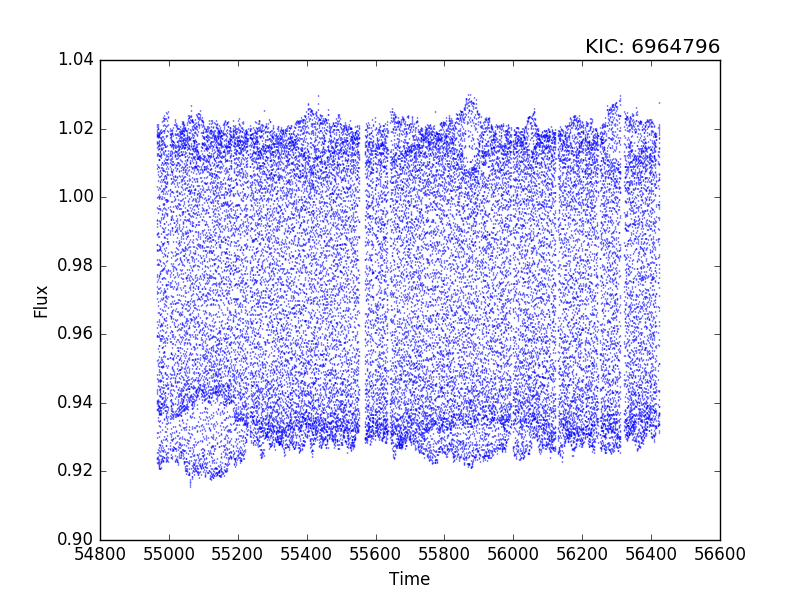} &
            \includegraphics[width=0.5\textwidth]{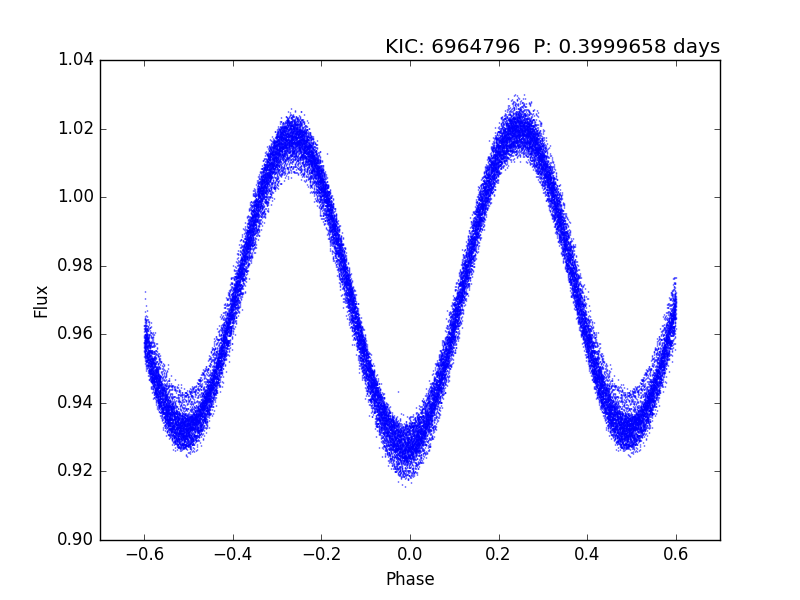} \\
            \includegraphics[width=0.5\textwidth]{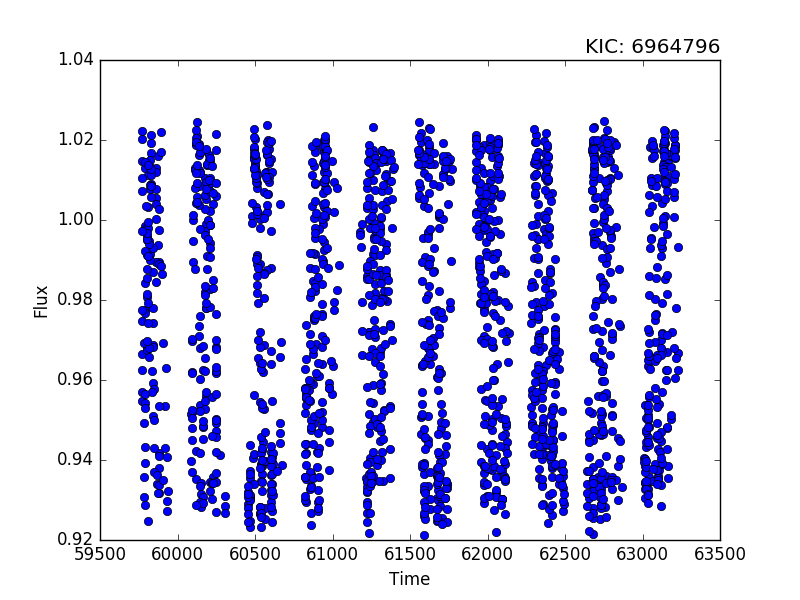} &
            \includegraphics[width=0.5\textwidth]{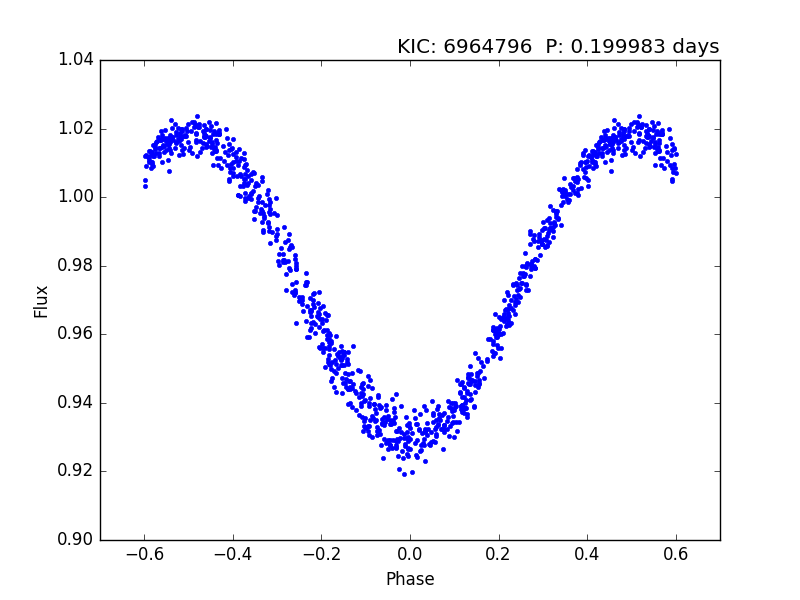}
        \end{tabular}
        \caption{
            \textbf{Top Left}: Original \kepler{} light curve.
            \textbf{Top Right}: The light curve phase-folded over the \kebc{} period.
            \textbf{Bottom Left}: \lsst{} simulated light curve.
            \textbf{Bottom Right}: The light curve phase-folded over the \aov{} period.
        }
        \label{fig:halfperfigs}
    \end{figure}

\begin{figure}[!tbp]
        \centering
        \begin{tabular}{cc}
            \includegraphics[width=0.5\textwidth]{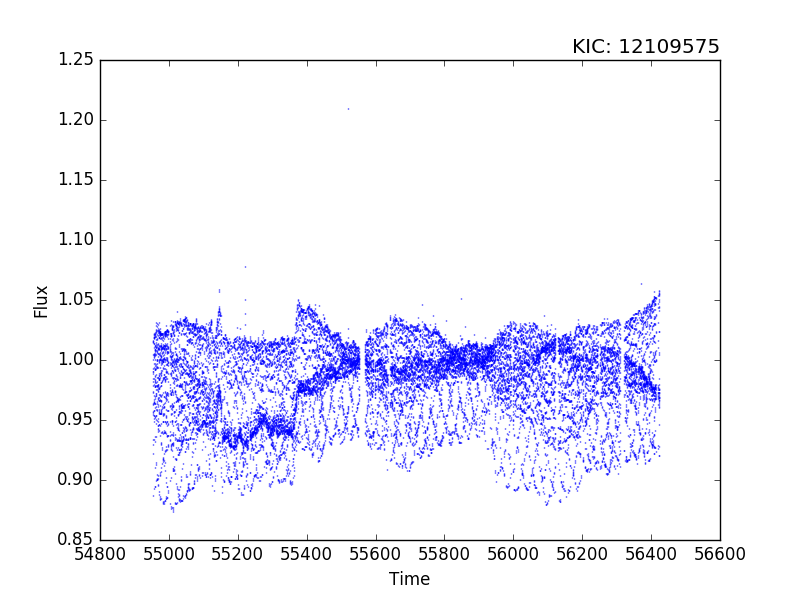} &
            \includegraphics[width=0.5\textwidth]{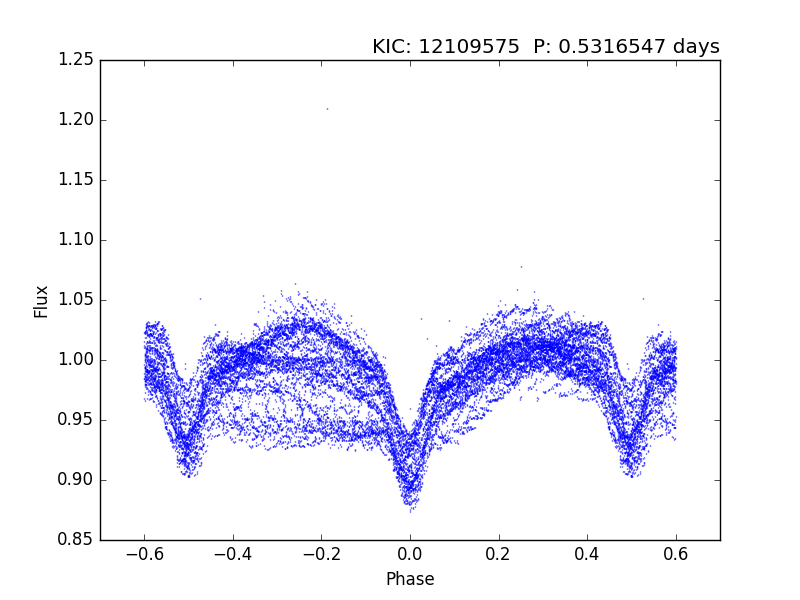} \\
            \includegraphics[width=0.5\textwidth]{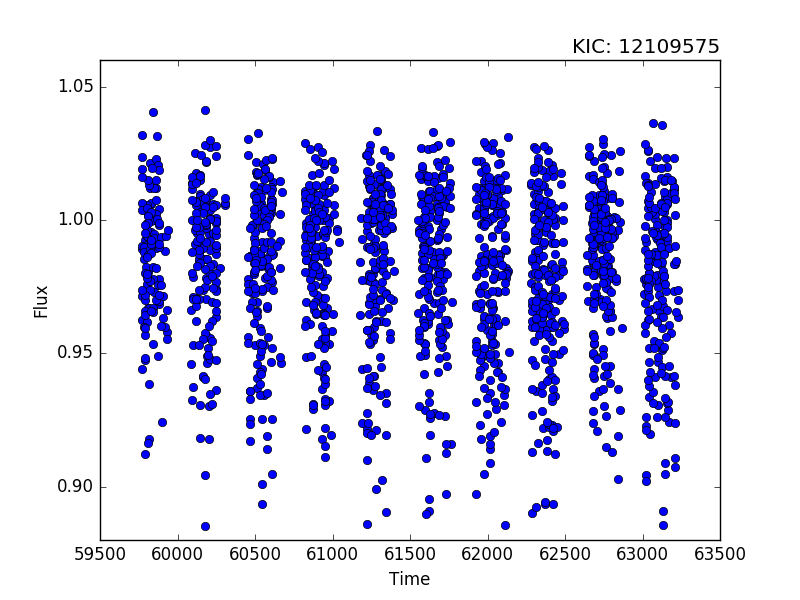} &
            \includegraphics[width=0.5\textwidth]{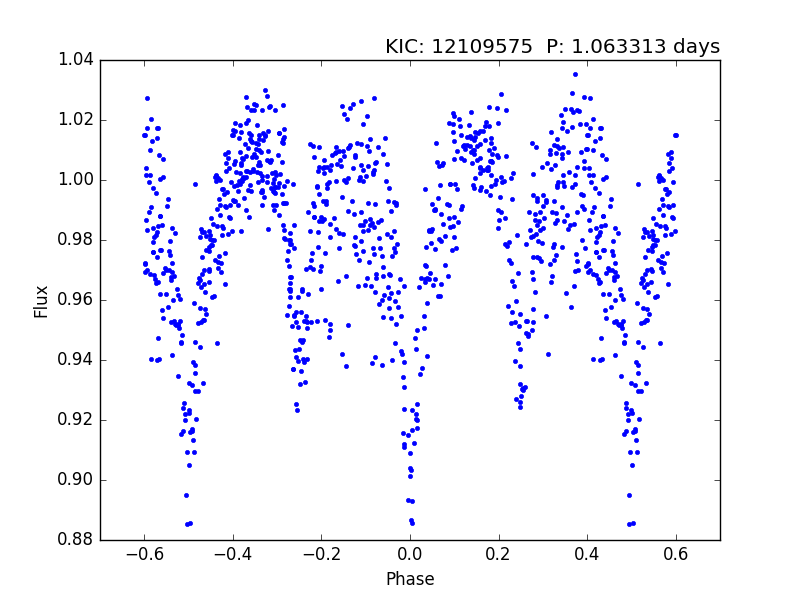}
        \end{tabular}
        \caption{
            \textbf{Top Left}: Original \kepler{} light curve.
            \textbf{Top Right}: The light curve phase-folded over the \kebc{} period.
            \textbf{Bottom Left}: \lsst{} simulated light curve.
            \textbf{Bottom Right}: The light curve phase-folded over the \aov{} period.
        }
        \label{fig:twiceperfigs}
    \end{figure}

    We define a successful detection (Table~\ref{tab:sucdet}) if $\pratio{}$ is within a specified tolerance range of one of three target values: 0.5, 1, and 2.
    We determined that a tolerance of 0.1\% was sufficient as increasing the tolerance beyond this point did not alter the results significantly (Table~\ref{tab:perres}).
    The values of $\pratio{}$ were chosen because they correspond to different  light curve shapes.
    A $\pratio{}$ of 0.5 corresponds to the half cycle of the binary.
    This is the tendency if the primary and secondary eclipses are similar in shape and depth.
    Figure~\ref{fig:halfperfigs} shows an example of one such target.
    A $\pratio{}$ of 1 indicates that the period finder returned the close-to-true orbital period of the binary.
    In contrast to the previous case, binaries with eclipses that are different depths and shapes have a strong orbital period signal.
    Figure~\ref{fig:perfigs} depicts a source in which the original binary period was recovered.
    A $\pratio{}$ of 2 corresponds to twice the period of the binary.
    Spots can change the depths of the eclipses from orbit to orbit.
    This effect can produce light curves where the strongest period is a multiple of the orbital period of the binary.
    For this work we do not consider binaries with a $\pratio{}$ greater than 2 times the orbital period.
    Figure~\ref{fig:twiceperfigs} shows an example of a binary with a $\pratio$ of 2.

    \begin{table}[!tbp]
        \centering
        \begin{tabular}{|lcll|}
            \hline
            if & $0.5(1-tol/2) \le \pratio \le 0.5(1+tol/2)$ & then & half-period was recovered\\
            if & $1.0(1-tol/2) \le \pratio \le 1.0(1+tol/2)$ & then & period was recovered\\
            if & $2.0(1-tol/2) \le \pratio \le 2.0(1+tol/2)$ & then & twice-period was recovered\\
            \hline
        \end{tabular}
        \caption{
            A recovery is successful if its $\pratio{}$ is within some tolerance of the detection criterion.
            For example, a 1\% tolerance value would give ranges of size 0.005, 0.01, and 0.02 for $\pratio{}$ values of 0.5, 1.0, and 2.0, respectively.
            The criterion ranges would then be 0.49975 to 0.50025, 0.995 to 1.005, and 1.99 to 2.01, respectively.
        }
        \label{tab:sucdet}
    \end{table}
    

    The period finder successfully recovered 2027 out of the simulated 2844 binaries, 71.3\% of the sample.
    Breaking this up into their detection types (see Table~\ref{tab:sucdet}), we find that 1133 (39.8\%) were identified by their half cycle signal and 887 (31.3\%) by the true binary period.
    There were 7 (0.2\%) that were detected with a $\pratio{}$ of 2.
    These results are listed in Table~\ref{tab:perres}.
    \begin{table}[!tbp]
        \centering
        \begin{tabular}{|c|rr|rr|rr|rr|rr|rr|rr|}
\hline
 & \multicolumn{2}{c|}{0.01\%} & \multicolumn{2}{c|}{0.02\%} & \multicolumn{2}{c|}{0.05\%} & \multicolumn{2}{c|}{0.1\%} & \multicolumn{2}{c|}{0.2\%} & \multicolumn{2}{c|}{0.5\%} & \multicolumn{2}{c|}{1.0\%} \\
 $\pratio{}$ & num. & perc. & num. & perc. & num. & perc. & num. & perc. & num. & perc. & num. & perc. & num. & perc. \\
\hline
\multirow{2}{*}{0.5} & 1066          & 37.48          & 1120          & 39.38          & 1140          & 40.08          & 1142          & 40.15          & 1142          & 40.15          & 1142          & 40.15          & 1142          & 40.15 \\
                     & \textit{1009} & \textit{35.48} & \textit{1058} & \textit{37.20} & \textit{1074} & \textit{37.76} & \textit{1076} & \textit{37.83} & \textit{1076} & \textit{37.83} & \textit{1076} & \textit{37.83} & \textit{1076} & \textit{37.83} \\
\multirow{2}{*}{1.0} & 796           & 27.99          & 860           & 30.24          & 878           & 30.87          & 879           & 30.91          & 879           & 30.91          & 879           & 30.91          & 879           & 30.91 \\
                     & \textit{ 763} & \textit{26.83} & \textit{ 814} & \textit{28.62} & \textit{ 832} & \textit{29.25} & \textit{ 832} & \textit{29.25} & \textit{ 832} & \textit{29.25} & \textit{ 832} & \textit{29.25} & \textit{ 832} & \textit{29.25} \\
\multirow{2}{*}{2.0} & 7             & 0.25           & 8             & 0.28           & 8             & 0.28           & 8             & 0.28           & 8             & 0.28           & 8             & 0.28           & 8             & 0.28 \\
                     & \textit{  11} & \textit{ 0.39} & \textit{  11} & \textit{ 0.39} & \textit{  11} & \textit{ 0.39} & \textit{  11} & \textit{ 0.39} & \textit{  11} & \textit{ 0.39} & \textit{  11} & \textit{ 0.39} & \textit{  11} & \textit{ 0.39} \\
\hline
\multirow{2}{*}{success} & 1869          & 65.72          & 1988          & 69.90          & 2026          & 71.24          & 2029          & 71.34          & 2029          & 71.34          & 2029          & 71.34          & 2029          & 71.34 \\
                         & \textit{1783} & \textit{62.69} & \textit{1883} & \textit{66.21} & \textit{1917} & \textit{67.41} & \textit{1919} & \textit{67.48} & \textit{1919} & \textit{67.48} & \textit{1919} & \textit{67.48} & \textit{1919} & \textit{67.48} \\
\multirow{2}{*}{failure} & 975           & 34.28          & 856           & 30.10          & 818           & 28.76          & 815           & 28.66          & 815           & 28.66          & 815           & 28.66          & 815           & 28.66 \\
                         & \textit{1061} & \textit{37.31} & \textit{ 961} & \textit{33.79} & \textit{ 927} & \textit{32.59} & \textit{ 925} & \textit{32.52} & \textit{ 925} & \textit{32.52} & \textit{ 925} & \textit{32.52} & \textit{ 925} & \textit{32.52} \\
\hline
\end{tabular}

    	\caption{
            Results of the period detection using a variety of tolerance values.
            The italicized rows are the results for the light curves with the injected $5\sigma$ noise.
            A tolerance of 0.1\% is sufficient as further increasing the tolerance does not alter the detection rates.
        }
        \label{tab:perres}
    \end{table}


\newpage
\section{Discussion}
\begin{figure}[!tbp]
        \begin{tabular}{cc}
            \includegraphics[width=\textwidth]{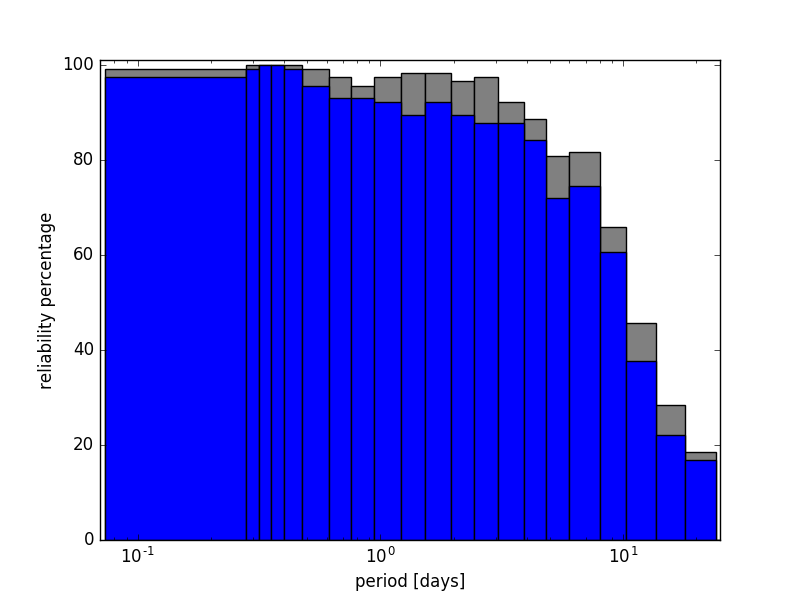}
        \end{tabular}
        \caption{
            The recovery rate, given as a fraction, as a function of the $\pkebc{}$.
            In blue are the results for the light curves with $5\sigma$ noise introduced while the results for the light curves with no additional noise are in gray.
            For periods less than 2~days we recover nearly 100\% of the binaries.
            The recovery success drops below 50\% for binaries with periods greater than 10~days.
        }
        \label{fig:successrate}
    \end{figure}

    Table~\ref{tab:perres} shows that 71.3\% of the binaries were recovered with the detection of the half-period being the most common.
    In addition, we also looked at how the binary period impacted detection success.
    Figure~\ref{fig:successrate} shows that, as expected, higher periods are detected less reliably.

    The recovery rate drops below 50\% for binaries with periods of above 10~days.
    It is also clear from the plot that the introduction of the $5\sigma$ noise suppresses the recovery rate.
    As expected, the suppression is period-dependent, because the geometric selection effect suppresses the detection of high eclipse amplitude light curves.
    The probability of an eclipse drops rapidly as the separation between components increases, which in turn means that we should expect suppression in high-amplitude light curves at the long period end.
    Because long period binaries will produce fewer eclipses as compared to short period binaries over the survey lifetime, the eclipse signal-to-noise ratio will be inherently lower for longer period binaries.

\begin{figure}[!tbp]
        \includegraphics[width=\textwidth]{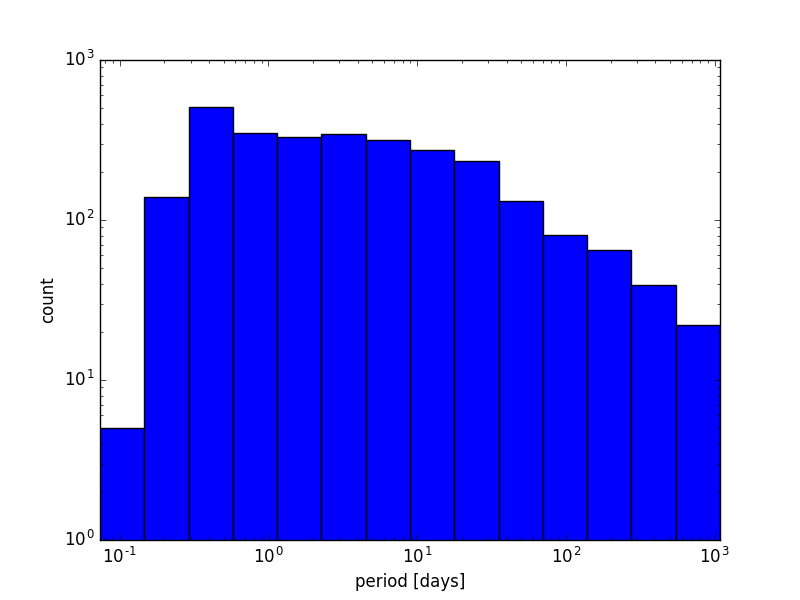}
        \caption{
            The number of binaries in the \kebc{} as a function of their period.
			The gradual drop off at higher periods is due to the geometrical probability of eclipses as well as \kepler{}'s duty cycle \citep{kebc3}.
        }
        \label{fig:kebcper}
    \end{figure}

    Figure~\ref{fig:kebcper} shows the number of binaries as a function of period in the \kebc{}.
    There are a substantial number of binaries to simulate even at periods of 100~days.
    It is not surprising that binaries with shorter periods will be detected more reliably.
    Shorter period binaries will have their eclipses more completely sampled, allowing for a more reliable detection.

    We explored what effect dithering would have on our results and there was no appreciable difference.
    Dithering refers to the intentional offset the scan pattern \lsst{} will make when returning to observe the same fields.
    In order to get full sky coverage, \lsst{} will have to have overlapping regions.
    Dithering will help smear the overlap so that coverage is more uniform.

\newpage
\section{Future Work}
    This work is a first step in the creation of a robust science metric for use in the determination of the \lsst{} cadence algorithm.
    Future steps include modifying the fluxes to be more representative of what \lsst{} would observe.
    Currently, the process is using \kepler{} fluxes without any modification.

    Using simulated instead of \kepler{} light curves and maintaining a standard number of light curves as a function of period would help remove potential biases due to \kepler{}'s selection effects.
    Longer period binaries suffer from an intrinsic bias due to the geometric probability of eclipse detection.

    Finally, constructing a light curve sample from a model of the Galaxy would allow us to evaluate the performance of \lsst{} as a function of position beyond just the \kepler{} field.
	This model would be generated using \galaxia{} \citep{Sharma2011} which implements the Besancon model \citep{Robin2003}.
    The model would also consider effects from crowding near the galactic plane as well as differences in stellar density, providing (within the accuracy of the Galaxy model) a more accurate representation of the observations \lsst{} will be making.

\acknowledgments
	Mark Wells and Andrej Pr\u{s}a acknowledge NSF grant \#1517460 that partly funded this research and ISSI project \#377~(2016).
	This research made use of Astropy, a community-developed core Python package for Astronomy \citep{astropy2013}.
    This research also made use of GNU Parallel \citep{Tange2011a}.

\bibliography{refs}{}
\end{document}